\documentclass[english]{article}
\usepackage[T1]{fontenc}
\usepackage[latin1]{inputenc}
\usepackage{amsmath}
\usepackage{amssymb}
\usepackage{esint}

\makeatletter

\newtheorem{theorem}{Theorem}[section]
\newtheorem{proposition}{Proposition}[section]
\newtheorem{lemma}{Lemma}[section]
\newtheorem{problem}{Problem}

\def\tire{\thinspace--\thinspace}
\def\R{{\mathchoice{{\bf R}}{{\bf R}}{{\rm R}}{{\rm R}}}}
\def\Z{{\mathchoice{{\bf Z}}{{\bf Z}}{{\rm Z}}{{\rm Z}}}}
\def\P{\mathop{\hbox{\sf P}}\nolimits}
\date{}

\makeatother

\begin{document}

\title{Fixed Points for Stochastic Open Chemical Systems}

\author{V.A.~Malyshev}
\maketitle
\begin{abstract}
In the first part of this paper we give a short review of the hierarchy
of stochastic models, related to physical chemistry. In the basement
of this hierarchy there are two models --- stochastic chemical kinetics
and the Kac model for Boltzman equation. Classical chemical kinetics
and chemical thermodynamics are obtained as some scaling limits in
the models, introduced below. In the second part of this paper we
specify some simple class of open chemical reaction systems, where
one can still prove the existence of attracting fixed points. For
example, Michaelis\tire Menten kinetics belongs to this class. At
the end we present a simplest possible model of the biological network.
It is a network of networks (of closed chemical reaction systems,
called compartments), so that the only source of nonreversibility
is the matter exchange (transport) with the environment and between
the compartments. 
\end{abstract}

\section{Introduction}

Relation between the existing (mathematical) physical theory and future
mathematical biology seems to be very intimate. For example, equilibrium
is a common state in physics but in biology equilibrium means death.
Biology should be deeply dynamical but this goal seems unreachable
in full extent: even in simplest physical situations the time consuming
complexity of any study of local dynamics is out of the present state
of art. Thus the only possibility would be to consider simpler dynamical
models (mean field etc.) but to go farther in their structure. The
obvious first step should have been related to chemical kinetics and
chemical thermodynamics. Here we present a review of these first results
and discuss what should be the second step.

In the first part of this paper we give a short review (in more general
terms than in \cite{Mal3}) of the hierarchy of stochastic models,
related to physical chemistry. In the basement of this hierarchy there
are two models --- stochastic chemical kinetics and the Kac model
for Boltzman equation. Classical chemical kinetics and chemical thermodynamics
are obtained as some scaling limits in the models, introduced below.

If some physical conditions, as reversibility, are assumed for a closed
(without matter exchange) system, then we have sufficiently simple
behaviour: one can prove convergence to a fixed point. However, in
many models of physical chemistry and biology, no reversibility condition
is assumed, and the behaviour can be as complicated as one can imagine.
Here we have already some gap between physics and biology, and it
is necessary to fill in this gap. In the second part of this paper
we specify some simple class of open chemical reaction systems, where
one can still prove the existence of attracting fixed points. For
example, Michaelis\tire Menten kinetics belongs to this class. At
the end we present a simplest possible model of the biological network.
It is a network of networks (of closed chemical reaction systems,
called compartments), so that the only source of nonreversibility
is the matter exchange (transport) with the environment and between
the compartments.

\section{Microdynamics}

Any molecule of mass $m$ can be characterized by translational degrees
of freedom (velocity $v\in\R^{3}$, coordinate $x\in\R^{3}$) and
internal, or chemical (for example, rotational and vibrational) degrees
of freedom. Internal degrees of freedom include the type $j=1,\ldots,J$
of the molecule and internal energy functionals $K_{j}(z_{j}),z_{j}\in\mathbf{K}_{j}$,
in the space $\mathbf{K}_{j}$ of internal degrees of freedom. It
is often assumed, see \cite{LanLif}, that the total energy of the
molecule $i$ is\[
E_{i}=T_{i}+K_{j}(z_{j,i}).\]
 We consider here the simplest choice when $K_{j}$ is the fixed nonnegative
number, depending only on $j$. It can be interpreted as the energy
of some chemical bonds.

We consider the set $\mathbf{X}$ of countable locally finite configurations
$X\!\!=\!\{x_{i},v_{i},j_{i}\}$ of particles (molecules) in $\R^{3}$,
where each particle $i$ has a coordinate $x_{i}$, velocity $v_{i}$
and type $j_{i}$. Denote by $\mathfrak{M}$ the system of all probability
measures on $\mathbf{X}$ with the following properties:
\begin{itemize}
\item Coordinates of these particles are distributed as the homogeneous
Poisson point field of particles on $\R^{3}$ with some density $c$. 
\item The vectors $(v_{i},j_{i})$ are independent of the space coordinates
and of the other particles. The velocity $v$ of a particle is assumed
to be uniformly distributed on the sphere with the radius defined
by the kinetic energy $T={m}(v_{1}^{2}+v_{2}^{2}+v_{3}^{2})/2$ of
the particle, and the pairs $(j_{i},T_{i})$ are distributed via some
common density $p(j,T),$ \[
\sum_{j}\int p(j,T)\, dT=1.\]
 
\end{itemize}
Our first goal will be to define random dynamics on $\mathbf{X}$
(or deterministic dynamics on $\mathfrak{M}$). It is defined by a
probability space $(\mathbf{X}^{0,\infty},\mu)$, where $\mu=\mu^{0,\infty}$
is a probability measure on the set $\mathbf{X}^{0,\infty}$ of countable
arrays $X^{0,\infty}(t)=\{x_{i}(t),v_{i}(t),j_{i}(t)\}$ of trajectories
$x_{i}(t),v_{i}(t),j_{i}(t)$ on intervals $I_{i}=(\tau_{i},\eta_{i})$,
where $0\leq\tau_{i}<\eta_{i}\leq\infty$. The measure $\mu$ belongs
to the set of measures $\mathfrak{M}^{0,\infty}$ on $X^{0,\infty}(t)$,
defined by the following properties:
\begin{itemize}
\item If for any fixed $0\leq t<\infty$ we denote by $\mu(t)$ the measure
induced by $\mu$ on $\mathbf{X}$, then $\mu(t)\in\mathbf{\mathfrak{M}}$. 
\item The trajectories $x_{i}(t),v_{i}(t),j_{i}(t)$ are independent, each
of them is a Markov process (not necessary time homogeneous). This
process is defined by initial measure $\mu(0)$ on $\mathbf{X}$,
by birth and death rates, defining time moments $\tau_{i},\eta_{i}$,
and by transition probabilities at time $t$, independent of the motion
of individual particles but depending on the concentration densities
$c_{t}(j,T)$ at time $t$. 
\item The evolution of the pair $(j,T)$ for the individual particle in-between
the birth and death moments is defined by the following Kolmogorov
equations, which control the one-particle process \begin{equation}
\frac{\partial p_{t}(j_{1},T_{1})}{\partial t}=\sum_{j}\!\int\!(P(t;j_{1},T_{1}|j,T)\, p_{t}(j,T)-P(t;j,T|j_{1},T_{1})p_{t}(j_{1},T_{1}))\, dT\label{kol_1}\end{equation}
 defining Markov process with distributions $p_{t}(j,T)$. The probability
kernel $P$ depends however on $p_{t}(j,T)$ itself, we shall make
it precise below. 
\end{itemize}
The dynamics we will describe here is based on some earlier mathematical
models and central dogmas of physical chemistry. The simplest way
to rigorously introduce the measure $\mu$ is by the limit of finite
volume random dynamics. Initial conditions for this dynamics are as
follows: at time $0$ some number $n^{(\Lambda)}(0)$ of molecules
are thrown uniformly in the cube $\Lambda$, their parameters $(j,T)$
are independent and have some common density $p_{0}(j,T)$, not depending
on $\Lambda$. Let $n_{j}(t)=n_{j}^{(\Lambda)}(t)$ be the number
of type $j$ molecules at time $t$.

\medskip{}

\noindent \underline{\sl Input-Output (I/O) processes}

\smallskip{}

Heuristically, our time scale is such that for the unit of time each
molecule does $O(1)$ transitions. Then for any macroquantity $q\textrm{ }$of
substance its $O(q)$ part may change. One should choose time scales
for input-output processes correspondingly.

The (output) rate of the jumps $n_{j}\rightarrow n_{j}-1$ is denoted
by $\lambda_{j}^{(0)}$, that is with this rate a molecule of type
$j$ is chosen randomly and deleted from $\Lambda$. Similarly, the
(input) rate of the jumps $n_{j}\rightarrow n_{j}+1$ is denoted by
$\lambda_{j}^{(i)}$, that is a molecule of type $j$ is put uniformly
in $\Lambda$ with this rate. Dependence of both rates on the concentrations
can be quite different. To get limiting I/O process after (canonical)
scaling one can assume that \begin{equation}
\lambda_{j}^{(0)}=f_{j}^{(0)}\Lambda,\qquad\lambda_{j}^{(i)}=f_{j}^{(i)}\Lambda,\label{IOfunc}\end{equation}
 where $f_{j},g_{j}>0$ are some functions of all $c_{1}^{(\Lambda)},\ldots,c_{J}^{(\Lambda)}$,
$c_{j}^{(\Lambda)}={n_{j}}/{\Lambda}$ are the concentrations. However,
mostly we restrict ourselves to the case when $f_{j}$ are functions
of $c_{j}$ only. In other words, an individual type $j$ molecule
leaves the volume with rate ${f_{j}^{(0)}(c_{j}^{(\Lambda)})}/{c_{j}^{(\Lambda)}}$.
Denote \begin{equation}
f_{j}=f_{j}^{(i)}-f_{j}^{(0)}.\label{IOfunc1}\end{equation}

\medskip{}

\noindent \underline{\sl Stochastic chemical kinetics$\vphantom{,}$}

\smallskip{}

The hierarchy presented here depends on what parameters of a molecule
are taken into account. In stochastic chemical kinetics only type
is taken into account. The state of the system is given by the vector
$(n_{1},\ldots,n_{J})$. There are also $R$ reaction types and the
reaction of the type $r=1,\ldots,R$, can be written as\[
\sum_{j}\nu_{jr}M_{j}=0\]
 where we denote by $M_{j}$ a $j$ type molecule, and the stoihiometric
coefficients $\nu_{jr}>0$ for the products and $\nu_{jr}<0$ for
the substrates. One event of type $r$ reaction corresponds to the
jump $n_{j}\rightarrow n_{j}+\nu_{jr},\; j=1,\ldots,J.$ Classical
polynomial expressions (most commonly used)\[
\lambda_{r}=A_{r}\prod\limits _{j:\nu_{jr}<0}n_{j}^{-\nu_{jr}}\]
 for the rates of these jumps define a continuous time Markov process,
a kind of random walk in $\Z_{+}^{J}$. This dependence can be heuristically
deduced from local microdynamics. However, polynomial dependence is
not the only possibility, see \cite{Sava}. Moreover, there can be
various scalings for these rates. The scaling\[
A_{r}=a_{r}\Lambda^{\gamma_{r}+1},\qquad\gamma_{r}=\sum_{j:\nu_{jr}<0}\nu_{jr}\]
 where $a_{r}$ are some constants and $\Lambda$ is some large parameter,
is called canonical because the classical chemical kinetics equations
\begin{equation}
\frac{dc_{j}(t)}{dt}=\sum_{r}R_{j,r}(\vec{c}(t))\label{cck}\end{equation}
 for the densities\[
c_{j}(t)=\lim_{\Lambda\rightarrow\infty}\Lambda^{-1}n_{j}^{(\Lambda)}(t)\]
 follow in the large $\Lambda$ limit, with some polynomials $R_{j,r}$
(see below and \cite{MaPiRy}).

\begin{problem} It is important to give (at least heuristic) local
probabilistic models to explain other than polynomial dependence and
scalings for the rates. For example for arbitrary homogeneous functions
as in \cite{Sava}. \end{problem}

It is assumed that at time $t=0$ as $\Lambda\rightarrow\infty,$
$\Lambda^{-1}n^{(\Lambda)}(0)\longrightarrow c(0).$

Chronologically, the first paper in stochastic chemical kinetics was
by Leontovich \cite{Leo}, which appeared from discussions with A.N.~Kolmogorov.
Other references see in \cite{GadLeeOth}. In 70s stochastic chemical
kinetics for small $R,J$ was studied intensively, see reviews \cite{McQua,Kal}.
At the same time the general techniques to get limiting equations
(\ref{cck}) appears in probability theory \cite{VenFre,EthKur}.
Now there are many experimental arguments in favor of introducing
stochasticity in chemical kinetics \cite{AdAr,ArRoAd,GadLeeOth}.

\medskip{}

\noindent \underline{\sl Stochastic energy redistribution}

\smallskip{}

In the classical Kac model \cite{Kac} the molecules $i=1,\ldots,N$
have the same type, but each molecule $i$ has a velocity $v_{i}$
or kinetic energy $T_{i}$. In collisions the velocities (or the kinetic
energies) change somehow. There is still continuing activity with
deeper results concerning the Kac model, in particular convergence
rate, see for example \cite{CaCaLo}.

One should merge Kac type models with stochastic chemical kinetics.
Then each molecule $i$ acquires a pair $(j_{i},T_{i})$ of parameters:
type $j$ and kinetic energy $T$. However this is not sufficient
to get energy redistribution. One should introduce also {}``chemical''
energy. As it is commonly accepted, the general idea is that the energy
of chemical bonds of a substrate molecule can be redistributed between
product molecules, part of the energy transforming into heat. To describe
this phenomena in well-defined terms we introduce fast and slow reactions.
Fast reactions do not touch chemical energy, that is types, but slow
reactions may change both kinetic and chemical energies, thus providing
energy redistribution between heat and chemical energy.

Examples of reactions:

1. All chemical reactions are assumed slow --- unary (unimolecular)
$A\rightarrow B$, binary $A+B\rightarrow C+D$, synthesis $A+B\rightarrow C$,
decay $C\rightarrow A+B$ etc. In any considered reaction the total
energy conservation is assumed, that is the sum of total energies
in the left side is equal to the sum of total energies in the right
side of the reaction equation.

2. Fast binary reactions of the type $A+B\rightarrow A+B$, which
correspond to elastic collisions and draw the system towards equilibrium.

3. Fast process of heat exchange with the environment, with reactions
of the type $A+B\rightarrow A+B$, but where one of the molecules
is an outside molecule. 

If there is no input and output, then the Markov jump process is the
following. Consider any subset $i_{1}<\ldots<i_{m(r)}$ of $m(r)=-\sum_{j:\nu_{jr}<0}\nu_{jr}$
substrate molecules for reaction of type $r$.

On the time interval $(t,t+dt)$ these molecules have a {}``collision''
with probability ${\Lambda^{-(m(r)-1)}}b_{r}\, dt$, where $b_{r}$
is some constant. Let the parameters of these molecules be $j_{k}=j(i_{k}),T_{k}=T(i_{k})$.
Denote \[
T=\sum_{i=1}^{m}T_{i},\qquad K=\sum_{i=1}^{m}K_{j_{i}}\]
 and $T^{\prime},K^{\prime}$ are defined similarly for the parameters
$j_{1}^{\prime},\ldots,j_{m}^{\prime},T_{1}^{\prime},\ldots,T_{m^{\prime}}^{\prime}$
of $m^{\prime}$ product molecules. The reaction occurs only if \begin{equation}
T+K-K^{\prime}\geq0\label{energycon}\end{equation}
 and then the energy parameters of the product particles at time $t+0$
have the distribution defined by some conditional density $P_{r}(T_{1}^{\prime},\ldots,T_{m^{\prime}-1}^{\prime}|T_{1},\ldots,T_{m})$
on the set $0\leq T_{1}^{\prime}+\ldots+T_{m^{\prime}-1}^{\prime}\leq T+K-K^{\prime}$.
By energy conservation then\[
T_{m^{\prime}}^{\prime}=T+K-K^{\prime}-\sum_{i=1}^{m^{\prime}-1}T_{i}^{\prime}.\]
 This defines a Markov process $M_{\bar{A}}(t)$ on the finite-dimensional
space (note that $T\in R_{+}$) \[
Q_{\bar{A}}=\bigcup_{(n_{1},\ldots,n_{J})}R_{+}^{n_{1}}\times\ldots\times R_{+}^{n_{J}}\]
 where the union is over all vectors $(n_{1},\ldots,n_{J})$ such
that for the array $\bar{A}=(A_{1},\ldots,A_{Q})$ of positive integers
and for any atom type $q=1,\ldots,Q,$ \[
\sum_{j}n_{j}a_{jq}=A_{q}\]
 where $a_{jq}$ is the number of atoms of type $q$ in the $j$ type
molecule. In other words, each atom type defines the conservation
law $A_{q}=\mathrm{const}$.

Now, using conditional densities $P_{r}$, we define the {}``one-particle''
transition kernel \begin{equation}
P(t;j_{1},T_{1}|j,T)=\sum_{r}P^{(r)}(t;j_{1},T_{1}|j,T),\label{kernel}\end{equation}
 that is the sum of terms $P^{(r)}$ corresponding to reactions $r$,
which we define for some reaction types. For unimolecular reactions
$j\rightarrow j_{1}$ the product kinetic energy $T_{1}$ is uniquely
defined, thus $P_{j\rightarrow j_{1}}$ is trivial and for some constants
$u_{jj_{1}},$ \[
P^{(j\rightarrow j_{1})}=u_{jj_{1}}\delta(T+K-K_{1}-T_{1}).\]
 For binary reactions $j,j'\rightarrow j_{1},j'_{1},$ \begin{align*}
P^{(j,j'\rightarrow j_{1},\, j'_{1})} & =\sum_{j^{\prime},\, j_{1}^{\prime}}\int dT^{\prime}dT_{1}^{\prime}b_{j,j'\rightarrow j_{1},\, j'_{1}}P_{j,j'\rightarrow j_{1},\, j'_{1}}(T_{1}|T,T^{\prime})\\
 & \quad\times c_{t}(j^{\prime},T^{\prime})\,\delta(T+K+T^{\prime}+K^{\prime}-K_{1}-T_{1}-K_{1}^{\prime}-T_{1}^{\prime}).\end{align*}
 In particular, for {}``fast'' collisions (which do not change type)
we have the same transition kernel but with $j=j_{1},j'=j'_{1}$.
We see that $P^{(j,j'\rightarrow j_{1},j'_{1})}$ depend on the concentrations
$c_{t}(j,T)=p_{t}(j,T)\, c(t).$ They are defined via the Boltzman
type equation \begin{align}
\frac{\partial c_{t}(j_{1},T_{1})}{\partial t} & =f_{j}(c_{j})+\sum_{j}\int\big(P(t;j_{1},T_{1}|j,T)\, c_{t}(j,T)\notag\\
 & \quad-P(t;j,T|j_{1},T_{1})\, c_{t}(j_{1},T_{1})\big)\, dT\label{bol_1}\end{align}
 which is similar to the Kolmogorov equation but includes also birth
and death terms.

All technicalities about the derivation of the limiting processes
see in Appendix of \cite{Mal3}.

\medskip{}
 \underline{\sl Space dynamics}

\noindent \smallskip{}

To get thermodynamics we need also volume, pressure etc. Thus it is
necessary to define space dynamics and also scaling limit.

In the jump process, defined above, each particle $i$ independently
of the others, in random time moments \[
\tau_{i}(\omega)<t_{1i}(\omega)<\ldots<t_{in}(\omega)<\ldots<\sigma_{i}(\omega)\]
 changes its type and kinetic energy (thus velocity). For each trajectory
$\omega$ of the jump process we define the local space dynamics as
follows. It does not change types, energies, velocities, but only
coordinates. If at jump moment $t$ of the trajectory $\omega$ the
particle acquires velocity $\vec{v}(\omega)=\vec{v}(t+0,\omega)$
and has coordinate $\vec{x}(t,\omega)$, then at time $t+s$ \begin{equation}
\vec{x}(t+s,\omega)=\vec{x}(t,\omega)+\vec{v}(\omega)s\label{trans}\end{equation}
 unless the next event (jump), concerning this particle, of the trajectory
$\omega$ occurs on the time interval $\left[t,t+s\right]$. We assume
periodic boundary conditions or elastic reflection from the boundary.
We denote this process by $X_{\Lambda}(t)$, the state space of this
process is the sequence of finite arrays $X_{i}=\left\{ j_{i},\vec{x}_{i},\vec{v}_{i}\right\} $.
Thus each particle $i$ has a piecewise linear trajectory in the time
interval $(\tau_{i},\sigma_{i})$.

\begin{theorem} The thermodynamic limit $X^{0,\infty}(t)=\mathfrak{X}_{c}(t)$
of the processes $X_{\Lambda}(t)$ exists and its distribution belongs
to $\mathfrak{M}^{0,\infty}$. \end{theorem}

\textbf{Proof} See \cite{Mal3}.

\section{Scaling limit}

Now we define more restricted (than $\mathfrak{M}$) manifolds of
probability measures on $\mathbf{X}$: the grand canonical ensemble
for a mixture of ideal gases with one important difference --- fast
degrees of freedom are gaussian and slow degrees of freedom are constants
$K_{j}$, depending only on $j$.

We consider a finite number $n_{j}$ of particles of types $j=1,\ldots,J$
in a finite volume $\Lambda$. Remind that for the ideal gas of the
$j$ type particles the grand partition function of the Gibbs distribution
is \begin{align*}
\Theta(j,\beta) & =\sum_{n_{j}=0}^{\infty}\frac{1}{n_{j}!}\bigg(\prod\limits _{i=1}^{n_{j}}\int_{\Lambda}\int_{\R^{3}}\int_{\mathbf{I}_{j}}d\vec{x}_{j,i}\, d\vec{v}_{j,i}\bigg)\exp\beta\bigg(n_{j}(\mu_{j}-K_{j})-\sum_{i=1}^{n_{j}}\frac{m_{j}v_{j,i}^{2}}{2}\bigg)\\
 & =\sum_{n_{j}=0}^{\infty}\frac{1}{n_{j}!}(\Lambda\lambda)^{n_{j}}\exp\beta(\mu_{j}-K_{j})n_{j}=\exp(\Lambda\lambda_{j}\exp\beta\hat{\mu}_{j})\end{align*}
 where \[
\lambda_{j}=\beta^{-{3}/{2}}\Bigl(\frac{2\pi}{m_{j}}\Bigr)^{{3}/{2}},\qquad\hat{\mu}_{j}=\mu_{j}-K_{j}.\]
 General mixture distribution of $J$ types is defined by the partition
function $\Theta=\prod_{j=1}^{J}\Theta(j,\beta)$. The limiting space
distribution of type $j$ particles is the Poisson distribution with
concentration $c_{j}$. We will need the formulas relating $c_{j}$
and $\mu_{j}$: \begin{align}
c_{j} & =\frac{\langle n_{j}\rangle_{\Lambda}}{\Lambda}=\beta^{-1}\frac{\partial\ln\Theta}{\partial\mu_{j}}=\lambda_{j}\exp\beta\hat{\mu}_{j},\notag\\
\mu_{j} & =\beta^{-1}\ln\Bigl(\frac{\langle n_{j}\rangle}{\Lambda}\lambda_{j}^{-1}\Bigr)=\mu_{j,0}+\beta^{-1}\ln c_{j}+K_{j},\end{align}
 where $\mu_{j,0}=-\beta^{-1}\ln\lambda_{j}$ is the so called standard
chemical potential, it corresponds to the unit concentration $c_{j}=1$.
We put $c=c_{1}+\ldots+c_{J}$.

We will need Gibbs free energy $G$ and the limiting Gibbs free energy
per unit volume\[
g=\lim_{\Lambda\rightarrow\infty}\frac{G}{\Lambda}=\sum\mu_{j}c_{j}.\]

Define by $\mathfrak{M}_{0}\subset\mathfrak{M}$ the set of all such
measures for any $\beta,\mu_{1},\ldots,\mu_{J}$, and by $\mathfrak{M}_{0,\beta}$
its subset with fixed $\beta$.

In the process defined above the kinetic energies are independent
but may have not $\chi^{2}$ distributions, that is the velocities
may not have Maxwell distribution. We force them to have it by specifying
some trend to equilibrium process (elastic collisions) and heat transfer
(elastic collisions with outside molecules) processes.

Assume that there is a family $M(a),0\leq a<\infty$, of distributions
$\mu_{a}$ on $R_{+}$ with the following property. Take two i.i.d.\
random variables $\xi_{1},\xi_{2}$ with the distribution $M(a)$.
Then their sum $\xi=\xi_{1}+\xi_{2}$ has distribution $M(2a)$. We
assume also that $a$ is the expectation of the distribution $M(a)$.
Denote $p(\xi_{1}|\xi)$ the conditional density of $\xi_{1}$ given
$\xi$, defined on the interval $[0,\xi]$. We put\[
P^{(f)}(T_{1}|T,T^{\prime})=p(T_{1}|T+T^{\prime})\]
 and of course $T_{1}^{\prime}=T+T^{\prime}-T_{1}$. Denote the corresponding
generator by $H_{N}^{(f)}$.

We model heat transfer similarly to the fast binary reactions, as
random {}``collision'' with outside molecules in an infinite bath,
which is kept at constant inverse temperature $\beta$. The energy
of each outside molecule is assumed to have $\chi^{2}$ distribution
with $3$ degrees of freedom and with parameter $\beta$. More exactly,
for each molecule $i$ there is a Poisson process with some rate $h$.
Denote by $t_{ik},k=1,2,\ldots,$ its jump moments, when it undergoes
collisions with outside molecules. At this moments the kinetic energy
$T$ of the molecule $i$ is transformed as follows. The new kinetic
energy $T_{1}$ after transformation is chosen correspondingly to
conditional density $p$ on the interval $[0,T+\xi_{ik}]$, where
$\xi_{ik}$ are i.i.d.\ random variables having $\chi^{2}$ distribution
with density $cx^{{1}/{2}}\exp(-\beta x)$. Denote the corresponding
conditional density by $P^{(\beta)}(T_{1}|T)$. In fact, this process
amounts to $N$ independent one-particle processes, denote the corresponding
generator $H_{N}^{(\beta)}$.

Thus we can write the generator as

\[
H=H(s_{f},s_{\beta})=H^{(r)}+s_{f}H^{(f)}+s_{\beta}H^{(\beta)}\]
 where $H^{(r)}$ corresponds to slow reactions and $s_{f},s_{\beta}$
are some large scaling factors, which eventually will tend to infinity.

We will force the kinetic energies to become $\chi^{2}$ using the
limit $s_{f}\rightarrow\infty$.

\begin{theorem} The limits in distribution \[
\mathfrak{C}_{c}(t)=\lim_{s_{f}\rightarrow\infty}\mathfrak{X}_{c}(t),\qquad\mathfrak{O}_{c,\beta}(t)=\lim_{s_{\beta}\rightarrow\infty}\mathfrak{C}_{c}(t)\]
 exist for any fixed $t$. Moreover, the manifold $\mathfrak{M}_{0}$
is invariant with respect to the process $\mathfrak{C}_{c}(t)$ for
any fixed rates $u,b,h$. The manifolds $\mathfrak{M}_{0,\beta}$
are invariant with respect to $\mathfrak{O}_{c,\beta}(t)$. \end{theorem}

Thus, in the process $\mathfrak{C}_{c}(t)$ the velocities have Maxwell
distribution at any time moment. For the process $\mathfrak{O}_{c,\beta}(t)$
moreover, at any time $t$ the inverse temperature is equal to $\beta$,
that is there is heat exchange with the environment. Our individual
molecules still undergo Markov process, but simplified. At the same
time, the macrovariables undergo deterministic evolution on $\mathfrak{M}_{0,\beta}$.

\medskip{}
 \underline{\sl Markov property --- chemical kinetics restoration }

\noindent \smallskip{}

Note that initially the jump rates depend on the energies. We show
that, after the scaling limit, the process restricted on the types
will also be Markov. We assume that there are only unary and binary
reactions but we do not need reversibility assumption here.

\begin{lemma} The process, projected on types, that is the process
$(n_{1}(t),\ldots,$ $n_{J}(t))$ is Markov. It is time homogeneous
for unary reaction system and time inhomogeneous in general. \end{lemma}

\textbf{Proof} \
Recall that the jump rates were assumed to have simplest energy dependence,
that is collisions occur independently of the energies, but reactions
occur only if energy condition (\ref{energycon}) is satisfied. Write
$g_{\beta}(r)=\P(\left|\xi\right|>r)$ for the $\chi^{2}$ random
variable $\xi$ with inverse temperature $\beta$.

Assume $K_{1}\leq\ldots\leq K_{J}$ and consider first the case of
unary reactions. It is easy to see that the process $\mathfrak{O}_{c,\beta}(t)$
can be reduced to the Markov chain on $\left\{ 1,\ldots,J\right\} $
with rates $v_{jj^{\prime}}=u_{jj^{\prime}}$ if $j\geq j^{\prime}$,
and $v_{jj'}=g_{\beta}(K_{j^{\prime}}-K_{j})u_{jj^{\prime}}$ if $j<j^{\prime}$.
We used here that the kinetic energy distribution is $\chi^{2}$ at
any time moment.

Similarly for the binary reaction $j,j^{\prime}\rightarrow j_{1},\, j_{1}^{\prime}$
we define the renormalized Markov transition rates as $c(j,j^{\prime}\rightarrow j_{1},\, j_{1}^{\prime})=b_{j,j^{\prime}\rightarrow j_{1},j_{1}^{\prime}}$
if $K_{j}+K_{j^{\prime}}\geq K_{j_{1}}+K_{j_{1}^{\prime}}$ and \[
c(j,j^{\prime}\rightarrow j_{1},j_{1}^{\prime})=b_{j,j^{\prime}\rightarrow j_{1},j_{1}^{\prime}}\P\{\left|\xi_{1}+\xi_{2}\right|>K_{j_{1}}+K_{j_{1}^{\prime}}-(K_{j}+K_{j^{\prime}})\}\]
 if $K_{j}+K_{j^{\prime}}<K_{j_{1}}+K_{j_{1}^{\prime}}$. Here $\xi_{i}$
are independent and $\chi^{2}$ with inverse temperature $\beta$.
It is crucial here the use of the scaling limit for fast reactions.

Thus, in the thermodynamic limit we get the equations without the
energies, that is the classical chemical kinetics \begin{equation}
\frac{dc_{j}(t)}{dt}=\sum_{r}R_{j,r}(\vec{c}(t))+f(c_{j}).\label{occk}\end{equation}

\underline{\sl Example: monotonicity of Gibbs free energy
for closed system with only unary} \underline{\sl reactions$\vphantom{,}$}

\noindent \smallskip{}

Assume now that the continuous time Markov chain on $\{1,\ldots,J\}$
with rates $u_{jj'}$ is irreducible. We say that this Markov chain
is compatible with the equilibrium conditions \begin{equation}
\mu_{1}=\ldots=\mu_{J}\label{equi}\end{equation}
 if its stationary probabilities $\pi_{j}$, or stationary concentrations
$c_{j,e}=\pi_{j}c$, satisfy the following conditions \[
\ln c_{1,e}+(\mu_{1,0}+K_{1})=\ldots=\ln c_{J,e}+(\mu_{J,0}+K_{J}).\]

\textbf{Remark} This compatibility condition should appear naturally
in local dynamics, but it is not clear how to deduce it in the mean
field dynamics. Note that reversibility is not a sufficient condition
for the compatibility condition.

To exhibit monotonicity for dynamics one needs special Lyapounov functions
in the space of distributions. For Markov chains this is the Markov
entropy with respect to stationary measure $\pi_{j},$ \[
S_{M}=\sum p_{j}\ln\frac{p_{j}}{\pi_{j}},\]
 see for example \cite{Ligg}.

Recall that the equilibrium function --- Gibbs free energy $g(t)$
--- undergoes deterministic evolution together with the parameters
$\mu_{j}$ or $c_{j}$. We will show that at any time moment it coincides
with the Markov entropy up to multiplicative and additive constants.

\begin{theorem} If the compatibility condition (\ref{equi}) holds,
then \begin{equation}
g(t)=\mu c+\frac{1}{\beta C}S_{M}(t)\label{GFE1}\end{equation}
 and monotone behaviour of the Gibbs free energy density follows.
\end{theorem}

\textbf{Proof} \ We have \begin{align}
g & =\lim_{\Lambda}\frac{G}{\Lambda}=\sum_{j}c_{j}\mu_{j}=\beta^{-1}\sum_{j}c_{j}\ln c_{j}+\sum_{j}c_{j}(\mu_{j,0}+K_{j})\label{free_1}\\
 & =\beta^{-1}\sum_{j}c_{j}\ln c_{j}+\sum_{j}c_{j}(\mu-\beta^{-1}\ln c_{j,e})\notag\\
 & =\mu c+\beta^{-1}\sum_{j}c_{j}\ln\frac{c_{j}}{c_{j,e}}\notag\end{align}
 where the first and the second equalities are the definitions, in
the third and the fourth equalities we used the formula \begin{align}
\mu_{j} & =\beta^{-1}\ln\Big(\frac{\langle n_{j}\rangle}{\Lambda}\lambda_{j}^{-1}\Big)=\mu_{j,0}+\beta^{-1}\ln c_{j}+K_{j},\intertext{where}\mu_{j,0} & =-\beta^{-1}\ln\lambda_{j}=-\beta^{-1}\Big(-\frac{d_{j}}{2}\ln\beta+\ln B_{j}\Big)\label{standard}\end{align}
 is the so called standard chemical potential, it corresponds to the
unit concentration $c_{j}=1$ for the equilibrium density, see for
example \cite{Mal3}.

At the same time\[
S_{M}=\sum p_{j}\ln\frac{p_{j}}{\pi_{j}}=C\sum c_{j}\ln\frac{c_{j}}{c_{j,e}}.\]

We see that for unary reactions one does not need reversibility assumption.

\underline{\sl Monotonicity of Gibbs free energy for closed
  system with binary reactions}

\noindent \smallskip{}

For binary reactions a similar result holds (we will not formulate
it formally). However, we do not have Markov evolution for the concentrations
anymore. Instead, we have the Boltzman equation for the concentrations,
that is the so called nonlinear Markov chain on $\{1,\ldots,J\}$.
Then, instead of the Markov entropy one should take the Boltzman entropy
with respect to some one-point distribution $p_{j}^{(0)}$ (see definitions
in \cite{MaPiRy})\[
S_{H}(t)=-\sum p_{j}(t)\ln\frac{p_{j}(t)}{p_{j}^{(0)}}\]
 which coincides with the Markov entropy for ordinary Markov chains.
For the monotonic behaviour of the Boltzman entropy, one should assume
reversibility or a more general condition --- unitarity, called local
equilibrium in \cite{MaPiRy}. Under this condition the monotonicity
of the Boltzman entropy was proved in \cite{MaPiRy}. We get the same
formula as (\ref{GFE1}) if we replace $S_{M}$ by $-S_{H}$.

Note that under these conditions $p_{j}(t)$ is a time inhomogeneous
Markov chain. In fact, in the long run, that is as $t\rightarrow\infty$,
the transition rates for one-particle inhomogeneous Markov chain,
in the vicinity of the fixed point, is asymptotically homogeneous.
This shows that binary case is asymptotically close to the unary case.

\section{Open thermodynamic compartments}

\noindent \underline{\sl Reversible and nonreversible processes}

\smallskip{}

Our systems in finite volume evolve via Markov dynamics. It is not
known when and how this dynamics could rigorously be deduced from
the local physical laws. However, there are many arguments that reversibility
is a necessary condition for this. Reversibility is a particular case
of the unitarity property of the scattering matrix of a collision
process. It was called local equilibrium condition in \cite{MaPiRy,FaMaPi}).

The reversibility gives strong corollaries for the scaling limits
--- 1)~Boltzman monotonicity and 2)~attractive fixed points. We
call chemical networks with properties 1) and 2) {\em thermodynamic
compartments}. Denote the class of such systems $\mathbf{T}$. These
systems are a little bit more general than the systems, corresponding
to the systems with local physical laws (in particular, having convergence
to equilibrium property). For example, any unimolecular reaction system
belongs to $\mathbf{T}$, because, as we saw above, the Markov entropy
is the Boltzman entropy here. However, biological systems obviously
are not of class $\mathbf{T}$. There are different ways to generalize
class $\mathbf{T}$ systems.

The first one is quite common: in chemical and biological systems
stochastic processes usually are not assumed to be reversible. However,
without the reversibility assumption the time evolution could be as
complicated as possible (periodic orbits, strange attractors etc.).
That has advantages --- one can adjust to real biological situations,
and disadvantages --- too many parameters, even arbitrary functions.
Normally, the rate functions $R_{j,r}$ can be rather arbitrarily
chosen, typical example where this methodology is distinctly pronounced
is \cite{CCCCNT}, connections with physics lost etc. In other words,
theory becomes meaningless when one can adjust it to any situation.

Another way could be a hierarchy of procedures to introduce nonreversibility
in a more cautious way. Each further step to introduce nonreversibility
is as simple as possible and each is related to time scaling, for
example, reversible dynamics is time scaled and projected on a subsystem.
We start to study here the simplest type of such procedures. In our
case the Markov generator will be the sum of two terms, \begin{equation}
H=H_{\mathit{rev}}+H_{\mathit{nonrev}},\label{rev-nonrev}\end{equation}
 where the first one is reversible and the other one is not, but the
latter corresponds only to input and output processes. One of technical
reasons to choose such nonreversible hamiltonian is to keep invariance
of the manifolds $\mathfrak{M},\mathfrak{M}_{0},\mathfrak{M}_{0,\beta}$.

In principle, another philosophy is possible --- large deviation or
other rare event conditioning, this we do not discuss here.

\medskip{}
 \underline{\sl Example {\rm 1:} steady states for open unimolecular systems}

\noindent \smallskip{}

We consider the case with $J=2$ and unary reactions only, however
the following assertions help to understand how more general open
systems can behave. Consider first the thermodynamic limit, and then
the stochastic finite volume problem.

In the thermodynamic limit the following equations for the concentrations
$c_{j}(t),j=1,2$, hold: 
\[
\frac{dc_{1}}{dt}=-\nu_{1}c_{1}+\nu_{2}c_{2}+f_{1},\qquad\frac{dc_{2}}{dt}=\nu_{1}c_{1}-\nu_{2}c_{2}+f_{2},\]
where $\nu_{1}=u_{12},\;\nu_{2}=u_{21}$ and $f_{j}$ are defined
by (\ref{IOfunc1}). Possible positive (i.e., $c_{1},c_{2}>0$) fixed
points satisfy the following system: 
\[
f_{1}(c_{1})+f_{2}(c_{2})=0,\qquad-\nu_{1}c_{1}+\nu_{2}c_{2}+f_{1}(c_{1})=0.\]

For example, for constant $f_{j}$ a positive fixed point exists for
any $c$ sufficiently large and equals\[
c_{1}=\frac{\nu_{2}c-f_{2}}{\nu_{1}+\nu_{2}},\qquad c_{2}=\frac{\nu_{1}c-f_{1}}{\nu_{1}+\nu_{2}}.\]
 In the linear case, that is for $f_{j}=a_{j}c_{j}$, for the existence
of a positive fixed point it is necessary and sufficient that $a_{j}$
have different signs and $|a_{j}|<\nu_{1}+\nu_{2}$. Then the positive
fixed point is unique and is defined by\[
c_{1}=\frac{\nu_{2}c}{\nu_{1}+\nu_{2}-a_{1}}.\]
 For faster than linear growth of $f_{j}$ fixed points cannot exist
for large $c$.

We see from these formulas that the equilibrium fixed point\[
c_{1}=\frac{\nu_{2}c}{\nu_{1}+\nu_{2}},\qquad c_{2}=\frac{\nu_{1}c}{\nu_{1}+\nu_{2}}\]
 (for the corresponding closed system) is slightly perturbed if $f_{j}$
(or $a_{j}$) are small. Moreover, the perturbed fixed point is still
attractive. This is true in more general situations as well.

Now consider the stochastic (finite volume) case.

\begin{proposition} Assume that $f_{j}$ are constants. In a finite
volume the process is ergodic if $\sum f_{j}<0$, transient if $\sum f_{j}>0$
and null recurrent if $\sum f_{j}=0$. \end{proposition}

\textbf{Proof} \
Note that the number of particles is conserved and the number of states
is finite if there is no I/O, otherwise the Markov chain is countable:
a random walk on $Z_{+}^{2}=\{(n_{1},n_{2}):n_{1}n_{2}\geq0\}$. There
are jumps $(n_{1},n_{2})\rightarrow(n_{1}-1,n_{2}+1)$ or $(n_{1},n_{2})\rightarrow(n_{1}+1,n_{2}-1)$
due to reactions, denote their rates $\nu_{1}n_{1},\nu_{2}n_{2}$
correspondingly. There are also jumps $(n_{1},n_{2})\rightarrow(n_{1}\pm1,n_{2}),(n_{1},n_{2})\rightarrow(n_{1},n_{2}\pm1)$
due to input-output with the parameters $a_{j}\Lambda$ and $b_{j}\Lambda$
correspondingly.

Transience and ergodicity can be obtained using Lyapounov function
$n_{1}+n_{2}$ and the results from \cite{FaMaMe}. To prove null
recurrence note that for sufficiently large $c$ the system should
be in the neighbourhood of the fixed point, which exists for $c$
sufficiently large. Thus one can also use the same Lyapounov function.

General conclusion is that only null recurrent case is interesting.
However, models with constant rates are too naive. It is reasonable
that there are regulation mechanisms which give more complex dependence
of $f_{j}$ on the rates. Unfortunately, there is no firm theoretical
basis to get exact dependence of reaction and I/O rates on the densities.

\medskip{}
 \underline{\sl Example {\rm 2:} stochastic Michaelis\tire Menten kinetics}

\noindent \smallskip{}

The generator for Michaelis\tire Menten kinetics is of type (\ref{rev-nonrev})
only in some approximation. This model has 4 types of molecules: $E$
(enzyme), $S$ (substrate), $P$ (product) and $ES$ (substrate-enzyme
complex). There are 3 reactions \[
E+S\rightarrow ES,\quad ES\rightarrow E+S,\quad ES\rightarrow E+P\]
 with the rates $k_{1}\Lambda^{-1}n_{E}n_{S},k_{-1}n_{ES},k_{2}n_{ES}$
correspondingly. We can also fix somehow the output rate for $P$
and input rate for $S$.

If $k_{2}=0$ then, as a zero'th approximation, we have a reversible
Markov chain. In fact, there are conservation laws \[
n_{E}+n_{ES}=m(E),\qquad n_{S}+n_{ES}=m(S)\]
 for some constants $m(E),m(S)$. Thus we will have random walk for
one variable, say $n_{ES}$, on the interval $[0,\mathrm{min}(m(E),m(S))]$,
with jumps $n_{ES}\rightarrow n_{ES}\pm1$. Such random walks are
always reversible. The stationary probabilities for this random walk
are concentrated around the fixed point of the limiting equations
of the classical kinetics \begin{equation}
\frac{dc_{ES}}{dt}=k_{1}c_{S}c_{E}-(k_{-1}+k_{2})c_{ES}\label{MM1}\end{equation}
 defined by\[
c_{ES}=\frac{c_{S}}{a+bc_{S}}\]
 for some constants $a,b$, defined by $m(E),m(S)$. If $k_{2}>0$
but small compared to $k_{1},k_{-1},$ then up to the first order
in $k_{2}$ we have the $P$ production speed\[
\frac{dc_{P}}{dt}=k_{2}c_{ES}=k_{2}\frac{c_{S}}{a+bc_{S}}.\]
 We could also look on this kinetics as on the simple random walk.
We have to introduce (arbitrarily) output rate for the product $P$
and adjust the input rate of $S$ so that the system becomes null-recurrent.
In fact, due to the conservation law $n_{E}+n_{ES}=m(E)$ we have
random walk on the half strip $\left\{ (n_{S},n_{ES})\right\} =Z_{+}\times(0,m(E))$.
The null-recurrence condition can be obtained using methods of \cite{FaMaMe},
we will not discuss this here.

\section{Network of thermodynamic compartments}

We call thermodynamic compartments, introduced above, {\em networks
of rank}~1. We saw that they have fixed points, and thermodynamics
plays the central role there. It can be some tightly dependent and/or
space localized system of chemical reactions.

Network of rank 2 consists of vertices $\alpha$ --- networks of rank
1, and directed edges, that is compartments are organized in a directed
graph. Directed edge from compartment $\alpha$ to compartment $\alpha^{\prime}$
means that there is a matter flow from $\alpha$ to $\alpha^{\prime}$.
Matter exchange between two compartments suggests some transport mechanism.
It is natural that there is a time delay between the moments of departure
from $\alpha$ and arrival to $\alpha^{\prime}$. The simplest probabilistic
model could be the following. Each $j$ type molecule leaves $\alpha$
for the destination $\alpha'$ with rates $f_{j,\alpha,\alpha'}$,
similar to defined in (\ref{IOfunc}), and after some random time
$\tau(j,\alpha,\alpha')$ arrives to $\alpha^{\prime}$. Times $\tau(j,\alpha,\alpha')$
are independent and their distribution depends only on $j,\alpha,\alpha^{\prime}$.
One can imagine that there is an effective distance $L(\alpha,\alpha^{\prime})$
between $\alpha$ and $\alpha^{\prime}$ and some transportation mechanism,
which defines effective speed to go through this distance. For example,
it can be transport through membrane, which can be represented as
a layer $\left[0,L\right]\times R^{2}$ of thickness $L$. During
time $\tau(j,\alpha,\alpha')$ the particle is absent from the network,
it has left $\alpha$ but has not yet arrived to $\alpha'$.

Denote by $c_{\alpha,j}$ the concentration of type $j$ molecules
in the compartment~$\alpha$. Limiting equations are \begin{align*}
\frac{dc_{\alpha',j}(t)}{dt} & =f_{\alpha',j}^{(i)}(c_{\alpha'}(t))-f_{\alpha',j}^{(0)}(c_{\alpha'}(t))+\sum_{\alpha}f_{j,\alpha,\alpha'}(c_{\alpha}(t-\tau(j,\alpha,\alpha'))\\
 & \quad-\sum_{\alpha}f_{j,\alpha',\alpha}(c_{\alpha'}(t))+\sum_{r}\nu_{\alpha',jr}R_{\alpha',r}(c_{\alpha'}(t))\end{align*}
 where $c_{\alpha}=(c_{\alpha,1},\dots c_{\alpha,J})$, $f_{\alpha,j}^{(i)}$
is the input rate to $\alpha$ from external environment, $f_{\alpha,j}^{(0)}$
is the output rate from $\alpha$ to the external environment. Note
that these equations are random due to random delay times $\tau$.
In the first approximation one can consider $\tau$ constant, however
random time delays seem very essential to restore randomness on the
time scale, higher than microscopic, in the otherwise deterministic
classical chemical kinetics.

Note that the above written equations follow from a similar microscopic
model --- we will not formally formulate it, because it is obvious
from our previous constructions: the corresponding manifold is $\times_{\alpha\in A}\mathfrak{M}_{\alpha}$,
where $A$ is the set of compartments, $\mathfrak{M}_{\alpha}$ is
the manifold for the compartment $\alpha$.

The following problems and phase transitions can be discussed in the
defined model on the rigorous basis (in progress):


1. The method of thermodynamic bounds in the thermodynamic networks,
defined in \cite{Mavr}. 

2. (Phase transitions due to transport rates.) Normal functioning
of the network can be close to the system $\left\{ c_{\alpha,j,e}\right\} $
of equilibrium fixed points in each compartment $\alpha$. Such situation
can be called homeostasis. Homeostatic regulation --- keeping the
system close to some system $\left\{ c_{\alpha,j,e}\right\} $. If
there is no transport, then the compartments are independent and the
fixed points inside them are pure thermodynamic. Under some transport
rates the fixed points change in a stable way, they smoothly depend
on the transport parameters. However, under some change of the transport
rates, the fixed points may change drastically: the system goes to
other basin of attraction. 

3. (Phase transition due to time desynchronization.) It is known now
that even a decease can be a consequence of timing errors. For a network
of rank~2, having for instance a cyclic topology (this is called
circuit in \cite{Tho}), assume that the input rates change periodically
or randomly in time. The question is: to what process the concentrations
converge and with what speed ? This time behaviour could be the next
step in the analysis of the structure of logical networks in the sense
of \cite{Tho}. 

\end{document}